\RequirePackage{booktabs}

\documentclass[pdflatex,sn-mathphys-ay]{sn-jnl}

\usepackage{graphicx}%
\usepackage{multirow}%
\usepackage{amsmath,amssymb,amsfonts}%
\usepackage{amsthm}%
\usepackage{mathrsfs}%
\usepackage[title]{appendix}%
\usepackage{xcolor}%
\usepackage{textcomp}%
\usepackage{manyfoot}%
\usepackage{booktabs}%
\usepackage{algorithm}%
\usepackage{algorithmicx}%
\usepackage{algpseudocode}%
\usepackage{listings}%
\usepackage{orcidlink}
\usepackage{deluxetable}

\theoremstyle{thmstyleone}%
\theoremstyle{thmstyletwo}%

\theoremstyle{thmstylethree}%

\defcitealias{Pike2023}{P23}

\begin{document}

\title[Characterizing hole trap production due to proton irradiation in germanium cross-strip detectors]{Characterizing hole trap production due to proton irradiation in germanium cross-strip detectors}


\author*[1]{\fnm{Sean N.} \sur{Pike}\orcidlink{0000-0002-8403-0041}}\email{snpike@ucsd.edu}

\author[1,2]{\fnm{Steven E.} \sur{Boggs}\orcidlink{0000-0001-9567-4224}}
\author[1]{\fnm{Gabriel} \sur{Brewster}\orcidlink{0009-0005-2936-8516}}
\author[1]{\fnm{Sophia E.} \sur{Haight}\orcidlink{0000-0003-3469-7072}}
\author[1]{\fnm{Jarred M.} \sur{Roberts}\orcidlink{0000-0002-7660-2740}}
\author[3]{\fnm{Albert Y.} \sur{Shih}\orcidlink{0000-0001-6874-2594}}
\author[4]{\fnm{Joanna} \sur{Szornel}\orcidlink{0009-0002-1776-5475}}
\author[2]{\fnm{John A.} \sur{Tomsick}\orcidlink{0000-0001-5506-9855}}
\author[2]{\fnm{Andreas} \sur{Zoglauer}\orcidlink{0000-0001-9067-3150}}

\affil[1]{\orgdiv{Department of Astronomy \& Astrophysics}, \orgname{University of California, San Diego}, \orgaddress{\street{9500 Gilman Drive}, \city{La Jolla}, \postcode{92093}, \state{CA}, \country{USA}}}

\affil[2]{\orgdiv{Space Sciences Laboratory}, \orgname{University of California, Berkeley}, \orgaddress{\street{7 Gauss Way}, \city{Berkeley}, \postcode{94720}, \state{CA}, \country{USA}}}

\affil[3]{\orgdiv{NASA Goddard Space Flight Center}, \orgaddress{\city{Greenbelt}, \postcode{20771}, \state{MD}, \country{USA}}}

\affil[4]{\orgdiv{Nuclear Science Division}, \orgname{Lawrence Berkeley National Laboratory}, \orgaddress{\city{Berkeley}, \postcode{94720}, \state{CA}, \country{USA}}}


\abstract{We present an investigation into the effects of high-energy proton damage on charge trapping in germanium cross-strip detectors, with the goal of accomplishing three important measurements. First, we calibrated and characterized the spectral resolution of a spare COSI-balloon detector in order to determine the effects of intrinsic trapping, finding that electron trapping due to impurities dominates over hole trapping in the undamaged detector. Second, we performed two rounds of proton irradiation of the detector in order to quantify, for the first time, the rate at which charge traps are produced by proton irradiation. We find that the product of the hole trap density and cross-sectional area, $[n\sigma]_\mathrm{h}$, follows a linear relationship with the proton fluence, $F_\mathrm{p}$, with a slope of $(5.4\pm0.4)\times10^{-11}\,\mathrm{cm/p^{+}}$. Third, by utilizing our measurements of physical trapping parameters, we performed calibrations which corrected for the effects of trapping and mitigated degradation to the spectral resolution of the detector.}




\maketitle

\section{Introduction}
\label{sec:introduction}
The soft gamma-ray band remains an understudied regime in astronomy due to the difficulty of achieving high sensitivity in the ``MeV gap" between about 100\,keV and 100\,MeV. NASA's Compton Spectrometer and Imager (COSI) Small Explorer (SMEX) mission will provide an opportunity to probe astrophysical sources in the energy range of 0.2--5\,MeV with unprecedented imaging and excellent spectral resolution as well as sensitivity to polarization \citep{COSI2023}. This will be achieved via the reconstruction of Compton scattering events across a $4\,\times\,4$ stack of cross-strip germanium detectors (GeDs), each with dimensions $\mathrm{8\,cm\times8\,cm\times1.5\,cm}$. By utilizing a perpendicular strip readout geometry, 3-dimensional spatial information can be inferred from differential drift times of the electron and hole charge clouds produced by each photon interaction in the detector, which drift in opposite directions due to a high voltage applied across each detector \citep{Amman2000,Bandstra2010}. Additionally, by reading out event information on both sides of each detector, the drift properties of both the liberated electrons and the corresponding positively-charged holes can be probed, allowing us to disentangle effects which act on the two types of charge carriers independently. In the work presented here, we discuss the effects of electron and hole trapping in particular.

Charge trapping occurs when drifting charge carriers encounter small regions of built-up charge of opposite polarity, impeding the motion of the charge carrier. The amount of energy deposited by the photon interaction which produced the clouds is inferred from the integrated current induced in conducting electrodes by the motion of the charge carriers across the bulk of the detector. This principle is known as the Shockley-Ramo Theorem \citep{Shockley1938,Ramo1939}. Therefore, trapping of charge of a given polarity leads to the incomplete measurement of the total amount of energy deposited. This effect varies depending on the total distance between the interaction site and the corresponding electrode \citep{Boggs2023}. 

Charge traps are produced by physical defects in the bulk of the detector, and the polarity of the traps differs depending on the type of defect: electrons are trapped by impurities in the detector producing regions of excess positive charge, and holes are trapped by defects in the germanium crystal lattice which accumulate an excess negative charge. The latter can be produced by the interaction of the germanium with energetic ($>1\,\mathrm{MeV}$) neutrons and protons, which dislodge atoms in the germanium crystal lattice, creating vacancies and filling interstitial sites. This is an important consideration for a space mission like COSI-SMEX, but the anticipated effects of radiation damage on the detectors are still uncertain. COSI-SMEX will be launched into an equatorial Low Earth Orbit (LEO) (targeting 530\,km altitude and $0^{\circ}$ inclination), which is one of the most benign space radiation environments. However, the actual radiation exposure for this orbit is not well predicted by standard radiation environment models and is sensitive to the exact altitude and inclination achieved after launch. If the resulting damage is left unaccounted for, the spectral performance of the instrument will degrade over time. 

Radiation damage effects in GeDs have been extensively studied for fast neutron induced damage due to predominantly neutron exposure in nuclear accelerator experiments \citep{Kraner1968}. Proton induced damage, which dominates in the space environment, has been studied to a much lesser extent \citep{Pehl1978,Koenen1995}. However, both types of exposure produce disordered regions in the crystal structure that act as charge traps, leading to incomplete charge collection and resulting in degraded spectral performance. Charge trapping is often characterized by a mean free drift length, $\lambda$, which can vary significantly for holes ($\lambda_\mathrm{h}$) and electrons ($\lambda_\mathrm{e}$). In the case where the drift velocity, $v_{d}$, of the charge carriers dominates over the thermal velocity, $v_{th}$, the fraction of charges still free after traveling a distance $d$ is $e^{-d/\lambda}$. In reality, the drift and thermal velocities are comparable for the typical operating conditions of GeDs, in which case the charge trapping must be modeled more completely by considering the actual density, $n$, and cross section, $\sigma$, of the traps encountered by a cloud of charge carriers traveling through the detector. In this case, the fraction of charges still free after traveling along a path length, $l$, due to both drift and thermal motion, is given by $e^{-l/n\sigma}$. Thus, electron and hole trapping may be parameterized by the trapping products, $[n\sigma]^{-1}_\mathrm{e,h}$. The electron trapping product is often assumed to be infinite (i.e., no electron trapping), though as we have shown in \cite{Pike2023} (hereafter P23), detectors may exhibit intrinsic electron trapping. In fact, electron trapping can dominate over hole trapping for undamaged detectors.

While we anticipate that the radiation damage due to protons will scale proportionally to the total proton fluence, $F_\mathrm{p}$, the dependence of charge trapping on $F_\mathrm{p}$ has never been characterized in the literature. There are two studies that have investigated the effects of proton radiation damage on GeDs, one using two planar GeDs \citep{Pehl1978}, and the other using both a reverse-electrode and a conventional-electrode coaxial GeD \citep{Koenen1995}. For both studies the detectors were held at their operational cryogenic temperatures, similar to the exposure conditions for COSI-SMEX in orbit. Both studies characterize the radiation damage in terms of spectral degradation, and do not explicitly provide a relation between the trapping parameters and $F_\mathrm{p}$. In this paper, we present experimental measurements of the effects of radiation damage on charge trapping in a spare GeD from the predecessor to COSI-SMEX, COSI-balloon. Our investigation includes, for the first time, a measurement of the relationship between the hole trapping product, $[n\sigma]^{-1}_\mathrm{h}$ and the proton fluence, $F_\mathrm{p}$. Furthermore, we demonstrate the potential for correcting for the damage via a second-order calibration procedure that we have already demonstrated for undamaged detectors \citepalias{Pike2023}.

\section{Methods and Calibrations}
\label{sec:methods}

\begin{figure}
    \centering
    \includegraphics[width=0.4\textwidth]{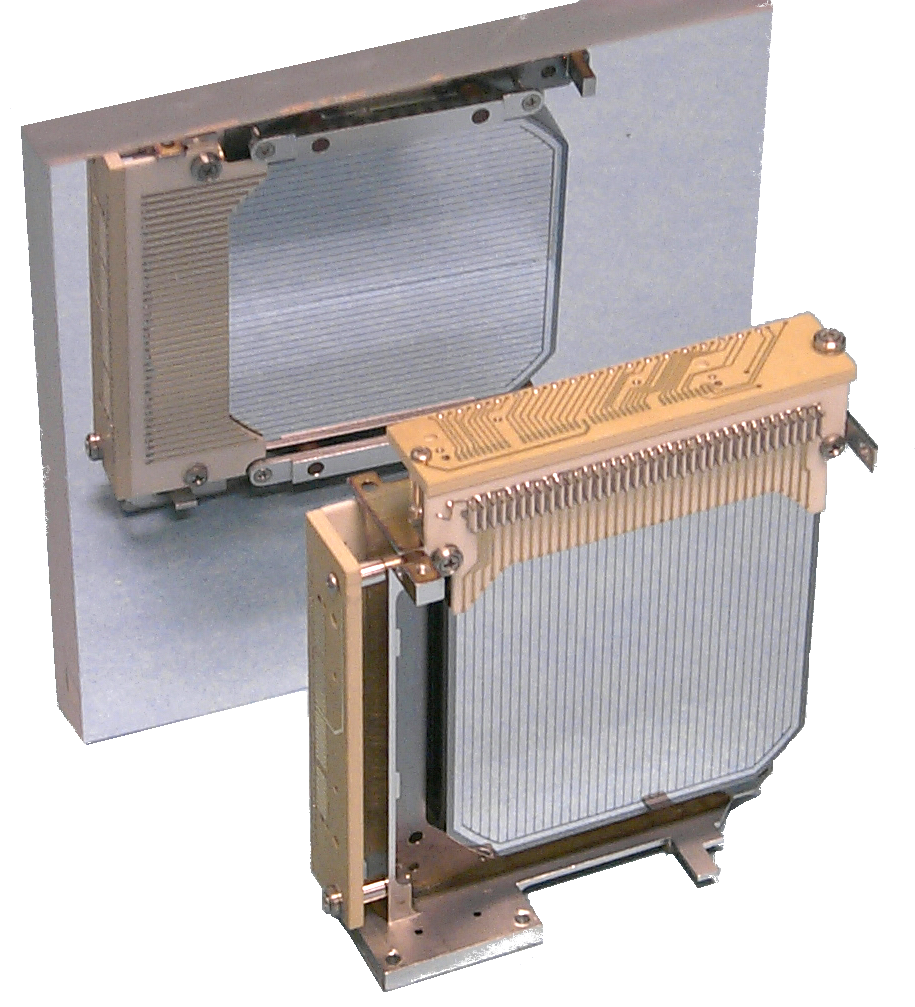}
    \caption{: A sample COSI-balloon GeD. Each detector consists of a high-purity germanium crystal with dimensions $\mathrm{8\,cm\times8\,cm\times1.5\,cm}$, instrumented with 37 orthogonal strip-geometry electrodes on the high-voltage and low-voltage sides. The orthogonal strips on the opposing face of the GeD are shown using a mirror.}
    \label{fig:det_pic}
\end{figure}

\begin{figure}
    \centering
    \includegraphics[height=1.34in]{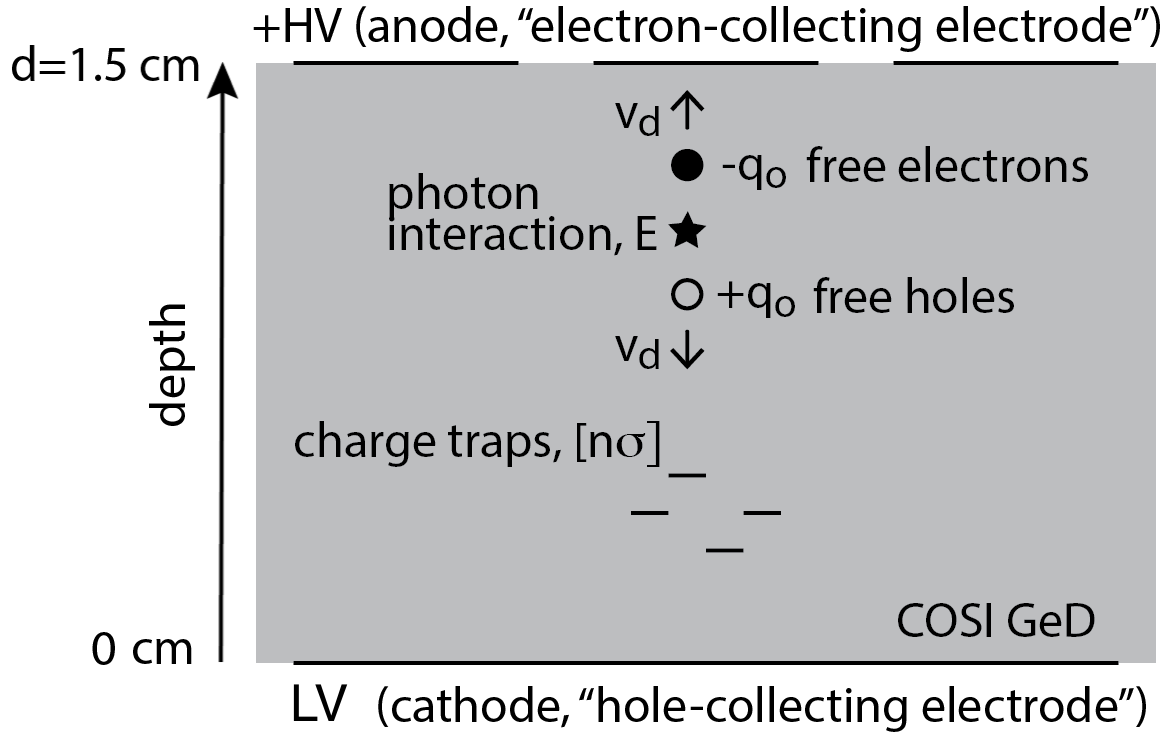}
    \includegraphics[height=1.34in]{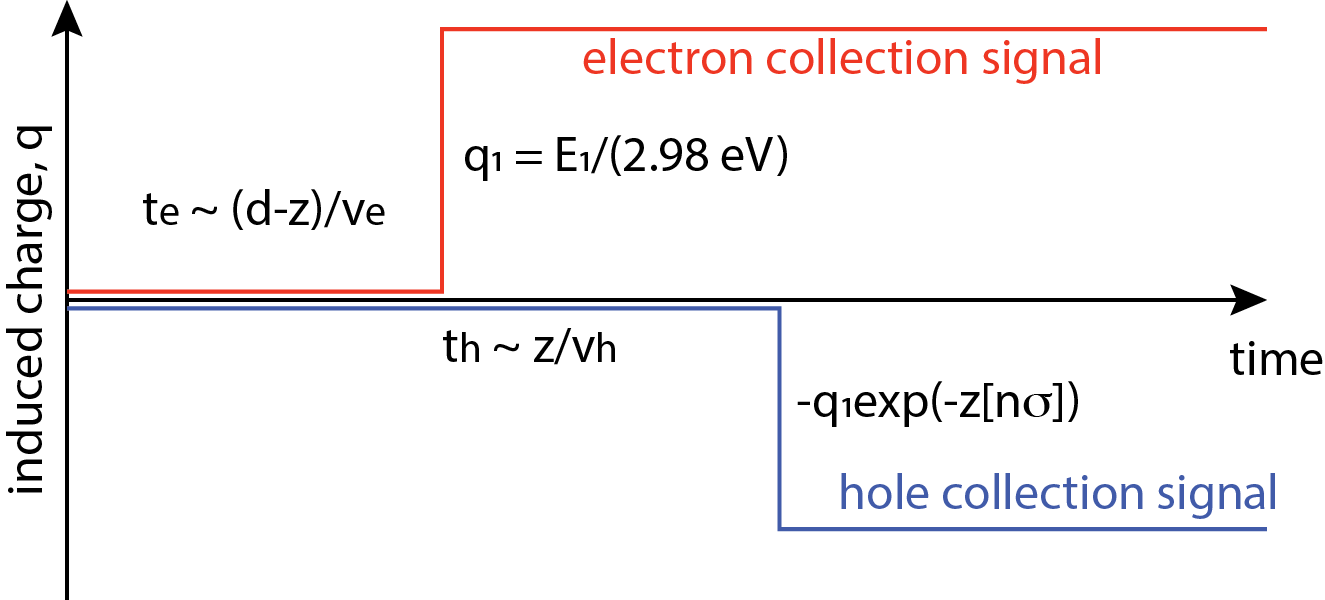}
    \caption{: Diagrams illustrating the relation between interaction depth and charge carrier drift length. Left: a diagram of the detector bulk. Right: a sketch of the charge induced on each side of the detector as a function of time. The further from a given collecting electrode that a photon interacts with the crystal, the longer it takes for the corresponding charge carriers to reach that electrode. The difference in drift time between carriers of opposite polarity allows for an estimate of the depth of photon interaction. Furthermore, the longer the drift length, the more charge traps a cloud of charge carriers will encounter, thereby decreasing the energy inferred from the motion of charge carriers.}
    \label{fig:depth_diagram}
\end{figure}

We used a spare COSI-balloon GeD with Detector ID HP41418-3 \citep{Kierans2018} to study the effects of damage-induced charge trapping. This detector has dimensions $8\,\mathrm{cm}\times8\,\mathrm{cm}\times1.5\,\mathrm{cm}$, and the high- and low-voltage (HV and LV, respectively) faces of the detector each have 37 strip-geometry electrodes, oriented perpendicularly in order to provide x- and y-dimension information for each photon interaction with a pixel size of about $2\,\mathrm{mm}\times2\,\mathrm{mm}$. Figure \ref{fig:det_pic} shows an example detector with the same dimensions and strip geometry. The impurity concentration of the detector is $2\times10^{9}\,\mathrm{cm^{-3}}$, and we operate the detector fully-depleted at a bias voltage of $+600$\,V. It is uncertain whether the detector is p- or n-type. We began by performing a baseline energy and depth calibration. We irradiated the undamaged detector with four radioactive sources, $^{241}\mathrm{Am}$, $^{133}\mathrm{Ba}$, $^{137}\mathrm{Cs}$, and $^{22}\mathrm{Na}$, in order to measure spectral lines with a range of energies between 60\,keV up to 1275\,keV. Using the \texttt{Melinator} and \texttt{Nuclearizer} software packages, both of which are described in \cite{Beechert2022} and which build upon the MEGAlib library \citep{Megalib}, we reconstructed the energy of each event as inferred independently by the HV (electron-collecting) and LV (hole-collecting) strips. We studied the effects of radiation damage on charge trapping by quantifying how the inferred energy of events shifted as a function of the z-position, or depth, of photon interaction. We determined the depth of each photon interaction using the difference in drift time (also referred to as the collection time difference, or CTD) between electrons and holes. Figure \ref{fig:depth_diagram} illustrates how depth is inferred from the CTD.

To determine how the measured CTD maps to the depth of photon interaction, we compared the CTD distributions observed for $^{241}\mathrm{Am}$ irradiation on the HV and LV sides to those determined via simulations. For each pixel, we determined a linear function that mapped the observed CTD values to the simulated values, and we interpolated the simulated relation between depth and CTD to finally arrive at the depth of each event. This method of depth-determination is adapted from the one described in Section 4.2 of \cite{Bandstra2010}. Throughout this paper, we define a depth of $z=0$ to be the face of the detector on which the LV electrode strips are located, and $z=1.5$\,cm is the face on which the HV electrode strips are located.

Next, we performed two rounds of proton irradiation. We first damaged the detector by bombarding it with a fluence of $2.00\times10^8\,\mathrm{p^{+}\,cm^{-2}}$ with an average proton energy of 150\,MeV. During the second round, we bombarded the detector with a fluence of $2.95\times10^{8}\,\mathrm{p^{+}\,cm^{-2}}$. The fluxes of the first and second irradiations were $2.23\times10^{5}\,\mathrm{p^{+}\,cm^{-2}\,s^{-1}}$ and $3.38\times10^{5}\,\mathrm{p^{+}\,cm^{-2}\,s^{-1}}$, respectively. Proton irradiation was performed at the James M. Slater, MD, Proton Treatment and Research Center at the Loma Linda University Medical Center. 
We hereafter refer to the time periods prior to radiation damage, after the first round of proton radiation, and after the second round of radiation damage as Epoch~I, Epoch~II, and Epoch~III, respectively. 

After each round of damage to the detector, we performed another round of calibration measurements using radioactive gamma-ray sources. During Epoch~II, we achieved a comparable total exposure time to the calibration campaign performed prior to proton irradiation. During Epoch~III, we only collected data using $^{241}\mathrm{Am}$ and $^{137}\mathrm{Cs}$. In order to directly compare the changes in the inferred photon energies before and after proton irradiation, we used the same energy calibration file (produced during Epoch I) to calibrate all datasets throughout the three epochs. We found that the CTD-to-depth relation changed slightly after damaging the detector, so we performed the depth calibration separately for the three epochs. 

Throughout the entire procedure discussed above, the detector was kept under vacuum (around $10^{-7}\,\mathrm{Torr}$) and at liquid nitrogen temperature (around 80\,K) in order to emulate the operating conditions of the COSI-SMEX detectors, which will be held at less than $85$\,K.

\section{Results}
\label{sec:results}

The goals of this study were threefold. First, we measured the intrinsic electron and hole trapping present in the high-purity germanium prior to proton damage. Next, we measured the effects of electron and hole trapping after the detector was damaged in order to determine how proton fluence impacts trapping. Finally, we performed second-order depth-dependent energy calibrations which corrected for the effects of trapping and mitigated degradation to the spectral resolution of the detector.

\subsection{Proton Fluence and Trapping Products}
\label{sec:trappingparams}

\begin{deluxetable}{cccccc}
    \tablenum{1}
    \tablecaption{Electron and hole trapping products determined by fitting simulated charge collection efficiency curves to centroid shifts as a function of depth. We show the results when electron trapping was allowed to vary between epochs, and the results when the electron trapping was fixed across epochs (denoted by a dagger, $\dagger$).\label{tab:trappingprods}}
    \tablewidth{0pt}
    \tablehead{\colhead{Epoch} & \colhead{$F_\mathrm{p}$} & \colhead{Lines fitted} & \colhead{$[n\sigma]^{-1}_{\mathrm{e}}$} & \colhead{$[n\sigma]^{-1}_{\mathrm{h}}$} & $\chi^2_{\nu}\ (\mathrm{d.o.f.})$ \\
      & \colhead{($\mathrm{p^{+}\,cm^{-2}}$)}  & & \colhead{(cm)} & \colhead{(cm)} & }
    \startdata
    I & $0$              & 2 & $734 \pm 7$  & $1790 \pm 30$ & $2.0\ (52)$ \\
    \hline
    \multirow{2}{*}{II}  & \multirow{2}{*}{$2\times10^{8}$}  & \multirow{2}{*}{2} & $728 \pm 17$  & $106.1 \pm 0.5$ & $24\ (52)$\\
     &   &   &  $734^{\dagger}$  & $106.2 \pm 0.5$   &   $24\ (53)$ \\
    \hline
    \multirow{2}{*}{III} & \multirow{2}{*}{$4.95\times10^{8}$}  & \multirow{2}{*}{1} & $510 \pm 26$  & $36.4 \pm 0.3$ & $134\ (52)$\\
     &   &   & $734^{\dagger}$   & $36.9 \pm 0.4$ & $223\ (53)$\\
    \enddata
    \vspace{-0.8cm}
    \tablecomments{The 1-sigma uncertainties listed for the trapping products were determined by fitting to the simulated CCE while multiplying the measured uncertainties of each depth plot by the square root of the corresponding reduced-$\chi^2$ value listed in the rightmost column. In this way, we attempt to propagate the systematic uncertainties in our CCE models.}
\end{deluxetable}

\begin{figure}[t!]
\centering
\includegraphics[width=0.49\textwidth]{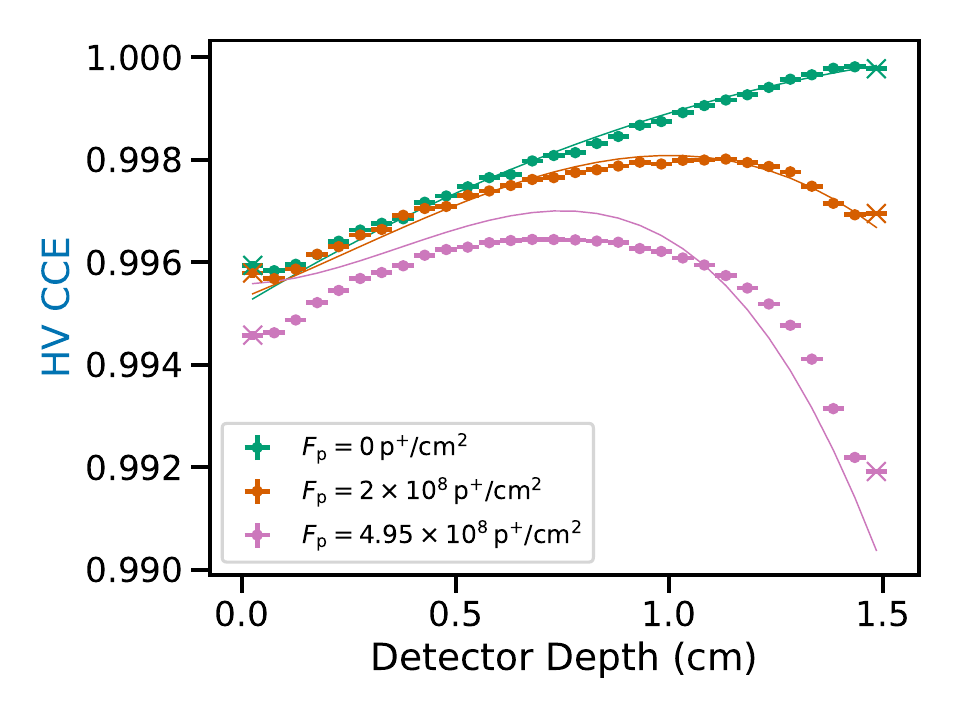}
\includegraphics[width=0.49\textwidth]{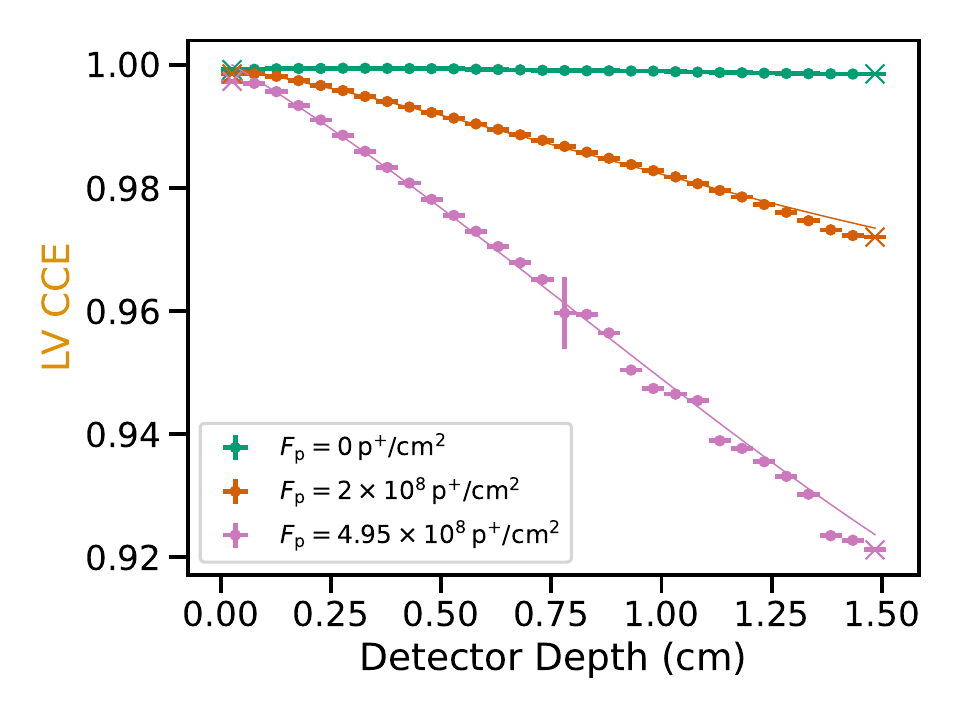}
\caption{: Normalized values of 662\,keV photopeak centroids measured by the HV (electron-dominated; left) and LV (hole-dominated; right) electrodes for different depth bins in the detector, fitted to simulated CCE curves shown as solid lines. The results are shown for Epochs I, II, and III (green, orange, and magenta, respectively). Points marked with an $\times$ were excluded during least-squares fitting to the simulated curves due to uncertainty in the depth calibration near the edges of the detector. 
The large error on the Epoch III LV data point around $0.75$\,cm is due to imperfect photopeak fitting for large amounts of trapping, which can result in suppression of the photopeak component compared to the low-energy tail components.}
\label{fig:CCE3epochs}
\end{figure}

For each radioactive source, we aggregated exposures into three datasets corresponding to each of the radiation epochs. We separated all single-pixel events (those which activated only one strip on each side of the detector) into 30 depth bins with a width of 0.5\,mm each. For each of these bins, we fit the spectral lines as measured by the HV and LV strips to the line profile model described in \citetalias{Pike2023}, which is composed of a Gaussian photopeak and two low-energy tails resulting from charge sharing between adjacent pixels. We recorded the line centroids as a function of detector depth and performed a least-squares fit of these ``depth plots" to the charge collection efficiency (CCE) curves given by Equation 8 in \cite{Boggs2023}:
\begin{equation}\label{eqn:CCE_energy}
    E(z) = A \times \left[1-B\times dCCE|_e(z)\right]\times\left[1-C\times dCCE|_h(z)\right]
\end{equation}
\noindent where $E(z)$ is the line centroid measured by a given electrode (HV or LV), and $A$ is a normalization factor which accounts for depth-independent shifts in the centroids introduced by the first-order gain calibration. The fractional centroid shift as a function of depth is given by $dCCE|_e(z)$ for the case when only electron trapping is simulated and $dCCE|_h(z)$ when only hole trapping is simulated. For each electrode polarity, we use one pair of simulated $dCCE|_{e,h}(z)$ templates which assume trapping products of $[n\sigma]_{e,h}^{-1}=1000$\,cm. The templates are scaled by the factors $B$ and $C$, which are inversely proportional to the electron and hole trapping products, respectively. Therefore, in this formalism, a value of $B=2$ ($C=2$) corresponds to an electron (hole) trapping product of $[n\sigma]^{-1}=500$\,cm.

We excluded the two data points closest to the faces of the detector from our least-squares fits because the precision with which the depth can be determined worsens near the electrodes. Although the electrodes on either side of the detector are most sensitive to charge carriers of opposite polarity, each signal is affected by both electron trapping and hole trapping. We therefore fit the HV and LV CCE curves simultaneously in order to constrain the electron and hole trapping products. Furthermore, for Epochs I and II, we simultaneously fit the HV and LV depth plots measured using the 356\,keV ($^{133}\mathrm{Ba}$) and 662\,keV ($^{137}\mathrm{Cs}$) photopeaks (resulting in 4 curves fit simultaneously) to a single pair of electron and hole trapping products, $[n\sigma]^{-1}_\mathrm{e,h}$. For Epoch III (following the second round of radiation damage), we did not collect $^{133}\mathrm{Ba}$ data, so we only fit the 662\,keV photopeak as a function of depth.
We did not collect sufficient data with $^{22}\mathrm{Na}$ to include the 1275\,keV photopeak in this analysis.
Additionally, we did not include the depth plots for the 60\,keV line emitted by $^{241}\mathrm{Am}$ in this analysis because the depth resolution of the detector is degraded at low photon energies, and because the line profile is complicated by Compton scattering in adjacent strips. 
We used the Python package \texttt{iMinuit} \citep{iminuit} to perform model fitting and error estimation for both spectral line fits and CCE fits.

The depth plots measured for the 662\,keV line emitted by $^{137}\mathrm{Cs}$ and the fitted CCE curves determined with the analysis described above are shown in Figure \ref{fig:CCE3epochs}. The trapping products which we determined are listed in Table \ref{tab:trappingprods} along with the reduced-$\chi^2$ statistic resulting from fitting the simulated CCE curves to the measured centroids as a function of depth. For Epochs II and III, we performed the fits twice: once with the electron trapping allowed to vary, and again with the electron trapping fixed to the value determined in Epoch~I, i.e. assuming that proton damage did not produce additional electron traps. The CCE curves shown in Figure \ref{fig:CCE3epochs} are those determined with fixed electron trapping.

We note that, despite the fact that the fits in the right panel of Figure \ref{fig:CCE3epochs} adequately reproduce the LV data, the fit statistic (determined via joint fitting of the HV and LV data) is generally not acceptable, and it becomes larger with increasing proton damage. Visual inspection of Figure \ref{fig:CCE3epochs} makes it clear that the model struggles to fit the HV (electron-dominated) data for Epoch~III. The HV CCE is more sensitive to the hole trapping products in Epochs II and III compared to Epoch I, and the fit statistic reflects that this is the case even when the electron trapping is allowed to vary. 

The increased fit statistic and the difficulty in reproducing the HV data may be attributed to multiple factors. The first is potential underestimation of the centroid errors in the depth plots. The errors which we input into the least-squares fit and those which we show in the depth plots in Figure \ref{fig:CCE3epochs} are those output by the \texttt{iMinuit} minimization routine. They do not take into account a detailed consideration of other sources of error, such as error in depth calibration. Second, our CCE model as well as our line profile model may diverge from reality for particularly large amounts of proton damage. We find that at the level of trapping present in Epoch III, even when binning events in depth, the line profiles exhibit increased width which produces degeneracy between the photopeak and low-energy tail components. As a result, line centroids may experience a systematic shift or increased error due to suppression of the photopeak amplitude in comparison to the amplitude of the low-energy tail components. The latter effect accounts for the particularly large error on the Epoch III data point at $0.75$\,cm in the right panel of Figure \ref{fig:CCE3epochs}. Similarly, at higher levels of hole trapping, uncertainties in the measured drift velocities, which were included in the CCE models, may become more apparent in the HV data. Furthermore, it is possible that is incorrect to assume that the drift speed remains constant regardless of trap density. This is an effect which we hope to investigate in future work. In order to account for these systematic uncertainties when determining the errors on the trapping products reported in Table \ref{tab:trappingprods}, we repeated the fits while multiplying the errors on the input data by the square root of the reduced-$\chi^2$ statistic. In this way, we effectively added error to the model in order to achieve a reduced-$\chi^2$ of unity, and this model error is propagated to the trapping products we report.

We find that this GeD, like the others we have studied, exhibits intrinsic charge trapping, even for holes. For the data collected in Epoch I, prior to proton damage, the simulated CCE curves provide a better fit to the depth plots relative to later epochs, with a reduced-$\chi^2$ of 2.0. We find that the intrinsic hole and electron trapping in this detector are comparable to the trapping products we reported in \cite{Boggs2023} for the COSI-balloon detector with ID HP41419-1 \citep{Kierans2018}. Here we observe somewhat smaller trapping products, indicating more intrinsic trapping. 

We also find that the electron trapping product did not change significantly between Epochs I and II. When the electron trapping product was allowed to vary, we obtained a value of $[n\sigma]^{-1}_\mathrm{e}=728\pm17\,\mathrm{cm}$ for Epoch~II. This value is consistent with the value obtained for Epoch~I, $[n\sigma]^{-1}_\mathrm{e}=734\pm7\,\mathrm{cm}$, and we therefore confirm that, at this level of irradiation, proton damage does not result in a significant increase in electron trapping. 

For Epoch~III, the model prefers more electron trapping, yielding a value of $[n\sigma]^{-1}_\mathrm{e}=510\pm26\,\mathrm{cm}$ when allowed to vary. However, as we have discussed, the model struggles to fit the HV data regardless of the value of electron and hole trapping. The electron trapping product may simply change to account for complicating effects on the electron-dominated signal, not considered in our model, which become apparent at high levels of hole trapping. We therefore do not consider this sufficient evidence for a physical change in the density of electron traps at this level of proton damage. For all epochs, the model is still able to reproduce the hole-dominated LV data with relatively good accuracy (see the right panel of Figure \ref{fig:CCE3epochs}). We therefore consider the hole trapping products obtained with these fits to be reflective of reality. This is further supported by the small errors we obtained for values of $[n\sigma]^{-1}_\mathrm{h}$ compared to $[n\sigma]^{-1}_\mathrm{e}$ for Epochs II and III, as well as the fact that regardless of whether the electron trapping is allowed to vary, we obtain consistent values for the hole trapping products.

\begin{figure}[t!]
\centering
\includegraphics[width=0.6\textwidth]{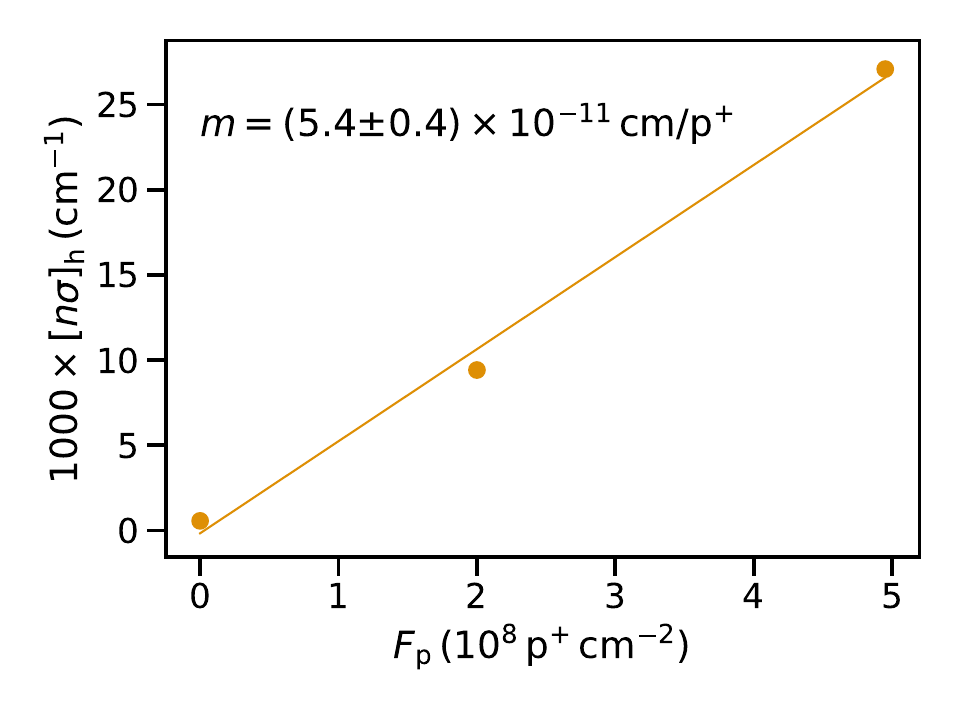}
\caption{: The reciprocal of the hole trapping product, determined by fitting measured centroid shifts as a function of depth to simulated CCE curves, is plotted as a function of proton fluence. We find a linear relation between the two parameters with slope, $m$.}
\label{fig:trapping_fluence}
\end{figure}

One result of our analysis is immediately apparent: damage due to proton irradiation increased hole trapping significantly. We have quantitatively characterized, for the first time, the relation between the hole trap density and the proton fluence applied to GeDs. In Figure \ref{fig:trapping_fluence}, we plot $[n\sigma]_\mathrm{h}$ as a function of proton fluence, $F_\mathrm{p}$, applied to the detector. Note that the y-axis is the reciprocal of the trapping product we have thus far referred to and which is listed in Table \ref{tab:trappingprods}. The reciprocal of the trapping product is directly proportional to the density of traps, clearly illustrating the increase in trapping as a function of proton fluence. We observed a roughly linear relation between $[n\sigma]_\mathrm{h}$ and the proton fluence, and we used Scipy \citep{SciPy} to calculate a linear regression. We obtained a slope of $m=(5.4\pm0.4)\times10^{-11}\,\mathrm{cm/p^{+}}$, representing the rate at which cross-sectional hole trapping area is generated by proton bombardment. We note that the proton flux differed between the two bombardments. While we do not observe clear evidence for flux-dependence of trap production, further investigation may be warranted in order to determine whether large differences in proton flux alter the rate of hole trap production.


\subsection{Trapping Corrections}
\label{sec:correction}

For Epochs I and II, we corrected the inferred photon energies associated with each event for the effects of trapping. This amounts to a depth-dependent second-order energy correction, similar to the one we presented in \citetalias{Pike2023}. We shifted the energy of each event according to the trapping products shown in Table \ref{tab:trappingprods}, using the fixed electron trapping result for Epoch II. For each event with depth $z$ and HV and LV energies $E_\mathrm{HV,LV}$, we calculated the corrected energy, $E^{\prime}_\mathrm{HV,LV} = E_\mathrm{HV,LV}/\mathrm{CCE_{HV,LV}}(z)$.

While for our previous study, we listed the full-width at half-maximum (FWHM) for the Gaussian component of a composite line function, here we list the FWHM and full-width at tenth-maximum (FWTM) determined via spline interpolation of the spectral lines. Due in large part to charge sharing between adjacent strips, this will lead to larger values than those listed in \citetalias{Pike2023}, but the values listed here more accurately capture the improvements obtained with this technique, as well as its limitations in the case of much higher hole trapping than we observed in \citetalias{Pike2023}.

\subsubsection{Before Proton Damage}
\label{sec:epoch1}

\begin{figure}[t!]
\centering
\includegraphics[width=5.0in]{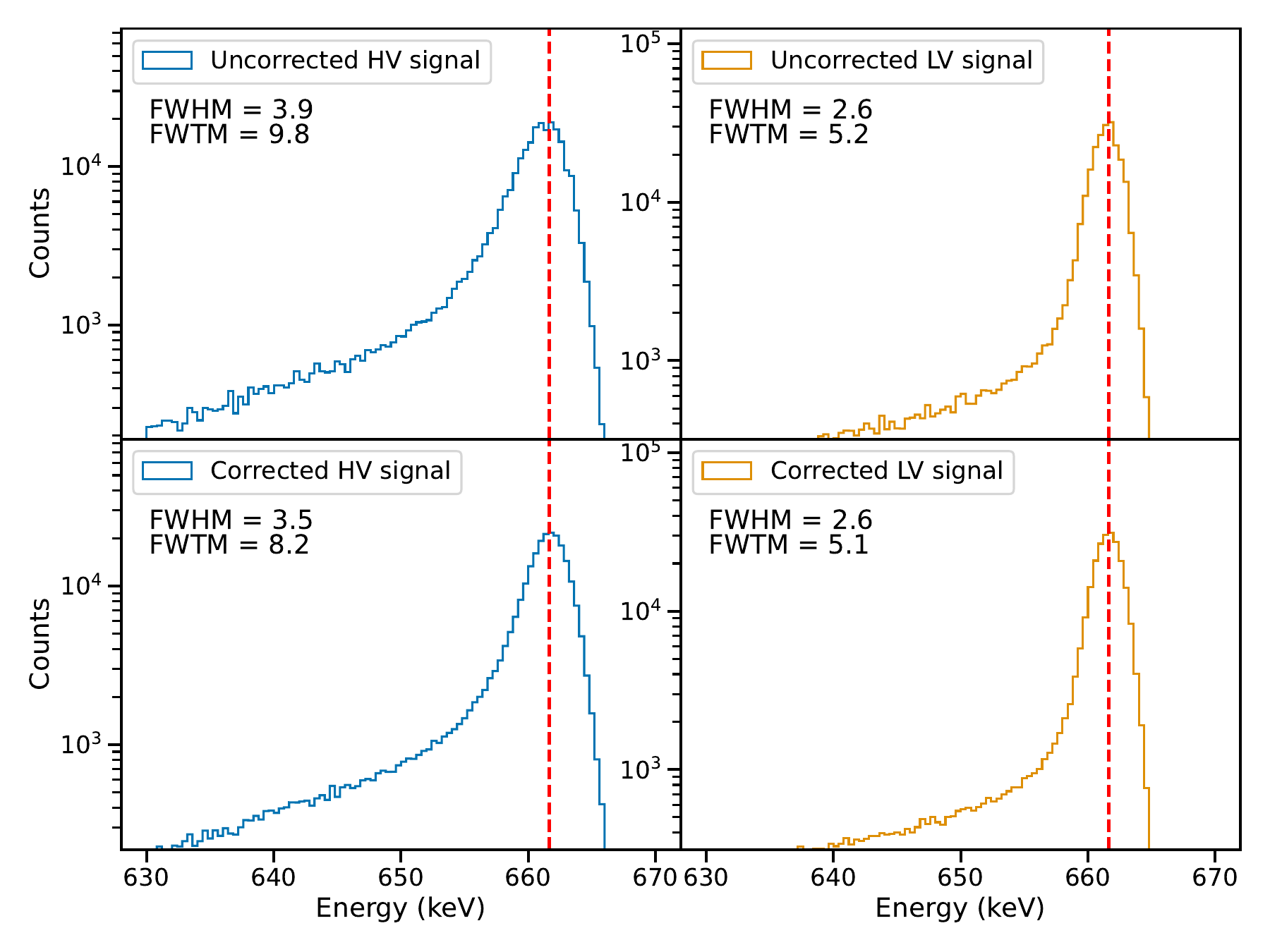}
\caption{: Measurements of the 662\,keV photopeak emitted from $^{137}\mathrm{Cs}$ using HV (blue) and LV (light orange) electrodes for Epoch I. The expected energy of the photopeak is shown as a red dashed line. The upper two panels show the spectral line measured before the depth-dependent trapping correction and the lower two panels show the corrected spectra. Before damage, electron trapping dominates. As a result, the trapping correction has a larger effect on the HV signal, significantly improving the low-energy tailing.}
\label{fig:Cs137specpreLL}
\end{figure}

\begin{deluxetable}{ccccc}
    \tablenum{2}
    \tablecaption{Values of the FWHM and FWTM of calibration lines measured with LV and HV electrodes during Epoch I (\textit{before} the detector was damaged). The FWHM and FWTM are given before and after trapping corrections were applied. Digits in parentheses represent the one-sigma error in the preceding digit. \label{tab:epoch1}}
    \tablewidth{0pt}
    \tablehead{ \colhead{Source} & \colhead{Photopeak Energy} & Electrode & \colhead{FWHM$_0$} & \colhead{FWHM$_\mathrm{corr.}$} \\
                    &        & &  \colhead{FWTM$_0$}  & \colhead{FWTM$_\mathrm{corr.}$} \\
                    & (keV)  & &  (keV)               &  (keV)    }
    \startdata
    \multirow{4}{*}{$\mathrm{^{241}Am}$}  & \multirow{4}{*}{60} & \multirow{2}{*}{LV}& $2.238(1)$  & $2.232(1)$    \\
                            &         & & $4.810(1)$  & $4.837(1)$  \\
                            \cline{3-5}
                            &         & \multirow{2}{*}{HV} & $2.576(1)$  & $2.588(1)$  \\
                            &         & & $6.676(1)$  & $6.578(1)$  \\
                            \hline
    \multirow{4}{*}{$\mathrm{^{133}Ba}$}   & \multirow{4}{*}{356}  & \multirow{2}{*}{LV}& $2.276(3)$  & $2.365(3)$  \\
                            &       & & $4.567(6)$  & $4.426(6)$  \\
                            \cline{3-5}
                            &         & \multirow{2}{*}{HV} & $3.172(4)$  & $2.937(4)$  \\
                            &         & & $7.11(1)$  & $6.47(1)$  \\
                            \hline
    \multirow{4}{*}{$\mathrm{^{137}Cs}$}   & \multirow{4}{*}{662}  & \multirow{2}{*}{LV}& $2.582(4)$  & $2.595(4)$   \\
                            &       & & $5.206(7)$  & $5.122(7)$  \\
                            \cline{3-5}
                            &         & \multirow{2}{*}{HV} & $3.912(5)$  & $3.505(5)$  \\
                            &         & & $9.81(1)$  & $8.20(1)$  \\
                            \hline
    \multirow{4}{*}{$\mathrm{^{22}Na}$}   & \multirow{4}{*}{1275}  & \multirow{2}{*}{LV}& $3.12(2)$  & $2.93(2)$  \\
                            &       & & $6.11(3)$  & $5.90(3)$   \\
                            \cline{3-5}
                            &         & \multirow{2}{*}{HV} & $5.31(3)$  & $4.69(3)$  \\
                            &         & & $16.97(9)$  & $13.65(7)$  \\\enddata
\end{deluxetable}

We list the FWHM and FWTM of each of the spectral lines to which we applied our correction technique for Epoch I in Table \ref{tab:epoch1}, and we show the corresponding corrected and uncorrected 662\,keV line measured using the HV and LV strips in Figure \ref{fig:Cs137specpreLL}. The top panels of Figure \ref{fig:Cs137specpreLL} show the spectra before we applied the trapping correction, and the bottom panels show the corrected spectra. When applying our correction to the spectra measured before the detector was damaged, we found that the spectral resolution of the HV strips was improved at all energies, with improvements to the FWTM being especially evident. Because intrinsic electron trapping dominates in the undamaged state (as illustrated in Figure \ref{fig:CCE3epochs}), the LV spectrum does not exhibit the same level of improvement as the HV spectrum, and for lower energies may even be slightly degraded due to the current limitations of the correction technique. However, we do observe a clear improvement in the FWTM obtained with the LV strips for all but the 60\,keV line, and we achieve an unambiguous improvement in both the HV and LV spectra at 1275\,keV.

\subsubsection{After Proton Damage}
\label{sec:epoch2}

\begin{figure}[t!]
\centering
\includegraphics[width=5in]{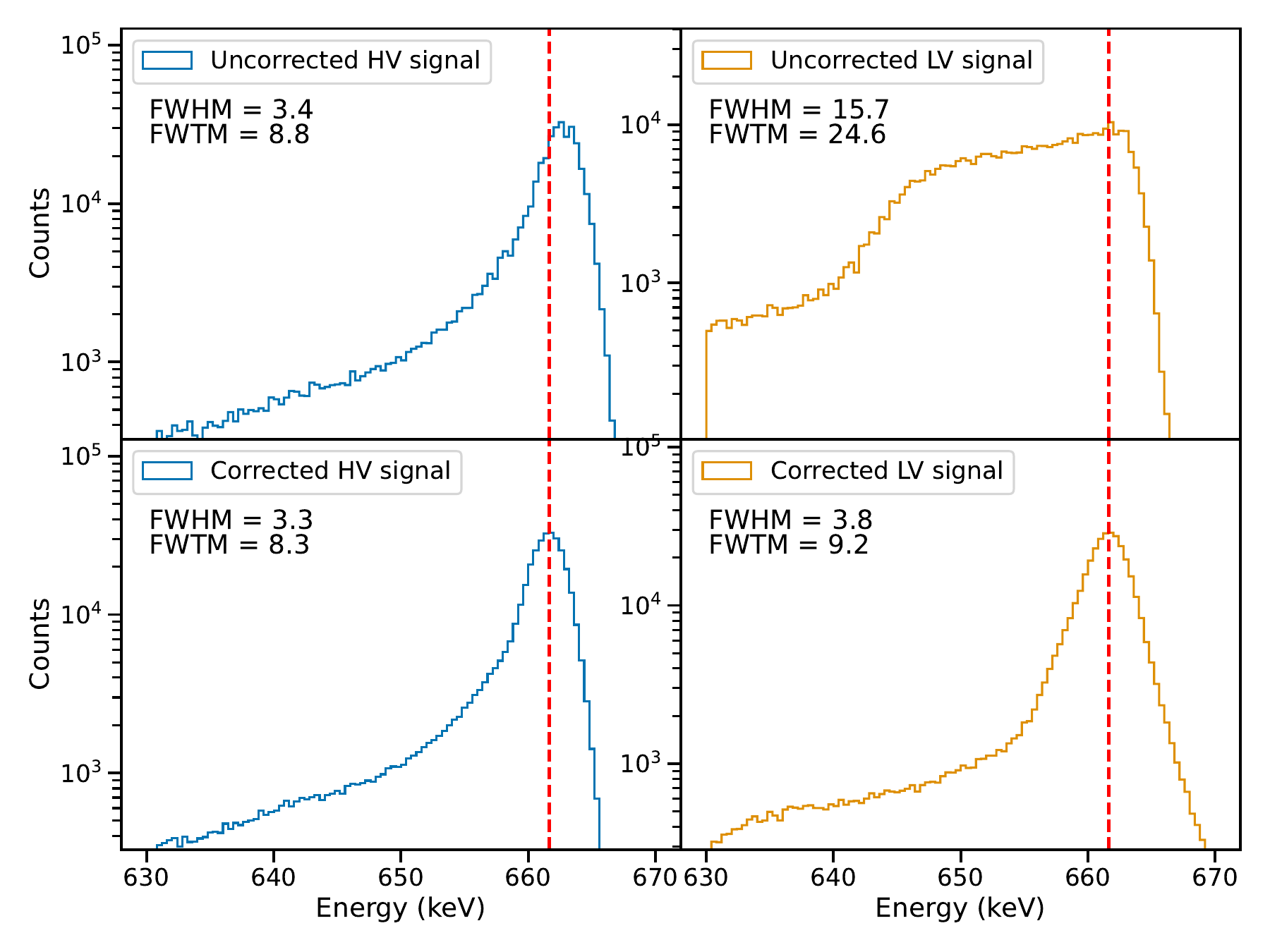}
\caption{: Measurements of the 662\,keV photopeak emitted from $^{137}\mathrm{Cs}$ using HV (blue) and LV (light orange) electrodes for Epoch II. The expected energy of the photopeak is shown as a red dashed line. The upper two panels show the spectral line measured before the depth-dependent trapping correction and the lower two panels show the corrected spectra. The effects of increased hole trapping are clear in both the HV and LV spectra. For the HV signal, the large increase in hole trapping actually narrows the line compared to the measurements taken before damage. The spectral line as measured by the LV electrodes experiences significant broadening, which we were largely able to correct.}
\label{fig:Cs137specpostLL1}
\end{figure}

\begin{deluxetable}{ccccc}
    \tablenum{3}
    \tablecaption{Values of the FWHM and FWTM of calibration lines measured with LV and HV electrodes during Epoch II (\textit{after} the detector was damaged). The FWHM and FWTM are given before and after trapping corrections were applied. \label{tab:epoch2}}
    \tablewidth{0pt}
    \tablehead{ \colhead{Source} & \colhead{Photopeak Energy} & Electrode & \colhead{FWHM$_0$} & \colhead{FWHM$_\mathrm{corr.}$} \\
                    &        & &  \colhead{FWTM$_0$}  & \colhead{FWTM$_\mathrm{corr.}$} \\
                    & (keV)  & &  (keV)               &  (keV)    }
    \startdata
    \multirow{4}{*}{$\mathrm{^{241}Am}$}  & \multirow{4}{*}{60} & \multirow{2}{*}{LV}& $3.051(1)$  & $2.263(1)$    \\
                            &         & & $6.707(1)$  & $4.921(1)$  \\
                            \cline{3-5}
                            &         & \multirow{2}{*}{HV} & $1.999(1)$  & $2.303(1)$  \\
                            &         & & $5.592(1)$  & $5.811(1)$  \\
                            \hline
    \multirow{4}{*}{$\mathrm{^{133}Ba}$}   & \multirow{4}{*}{356}  & \multirow{2}{*}{LV}& $9.70(1)$  & $2.943(3)$  \\
                            &       & & $15.06(1)$  & $6.247(5)$  \\
                            \cline{3-5}
                            &         & \multirow{2}{*}{HV} & $2.663(2)$  & $2.672(2)$  \\
                            &         & & $6.081(5)$  & $5.821(5)$  \\
                            \hline
    \multirow{4}{*}{$\mathrm{^{137}Cs}$}   & \multirow{4}{*}{662}  & \multirow{2}{*}{LV}& $15.67(2)$  & $3.816(4)$   \\
                            &       & & $24.65(3)$  & $9.20(1)$  \\
                            \cline{3-5}
                            &         & \multirow{2}{*}{HV} & $3.385(4)$  & $3.291(4)$  \\
                            &         & & $8.79(1)$  & $8.33(1)$  \\
                            \hline
    \multirow{4}{*}{$\mathrm{^{22}Na}$}   & \multirow{4}{*}{1275}  & \multirow{2}{*}{LV}& $26.6(1)$  & $5.53(2)$  \\
                            &       & & $57.3(2)$  & $14.57(5)$   \\
                            \cline{3-5}
                            &         & \multirow{2}{*}{HV} & $5.14(2)$  & $5.13(2)$  \\
                            &         & & $13.87(5)$  & $13.23(5)$  \\\enddata
\end{deluxetable}

We list the FWHM and FWTM of the spectral lines measured during Epoch II in Table \ref{tab:epoch2}, and we show the corresponding $^{137}\mathrm{Cs}$ spectra around 662\,keV in Figure \ref{fig:Cs137specpostLL1}. The uncorrected LV spectrum clearly illustrates the effects of large amounts of hole trapping on the spectral resolution of the detector. Before corrections, the spectrum measured by the LV strips exhibits dramatic broadening and low-energy tailing, resulting in a FHWM of $16$\,keV and a FWTM of $25$\,keV. These values are around four times larger than the FWHM and FWTM measured using the HV strips, which are significantly less sensitive to the increased hole trapping. 

We observe that the increased hole trapping has counteracted the effects of electron trapping on the HV signal for events near the HV strips, resulting in a narrower line than we observed before the detector was damaged. We also observe that the increased hole trapping coincides with a shift of the HV-measured energies towards higher values, essentially amounting to a change in the gain. This may or may not be causally linked with the increase in hole trapping, but may instead be a result of other changes in the experimental setup. For example, the detector temperature tends to drift slowly upward over time if the temperature is not cycled, perhaps due to moisture accumulation on the dewar-to-cryostat thermal coupling contact. Similarly, the temperature of the readout preamplifiers cannot be precisely controlled with our current experimental setup. Further testing of temperature dependence is necessary to disentangle experimental effects from the effects of proton damage.

The line width was improved significantly using our trapping correction procedure, bringing both the FWHM and the FWTM measured by the LV strips closer to those observed using the HV strips for all energies. However the line remains somewhat distorted, exhibiting a high-energy excess. The presence of this excess is an indicator that the correction we have applied is imperfect, which may be expected given finite precision in the determination of interaction depth. Nonetheless, the improvement in spectral resolution that we achieve using this correction technique is quite promising, with reductions in the FWHM of the LV signal amounting to 25-80\% depending on photopeak energy.

\section{Conclusions}
\label{sec:conclusions}

\begin{figure}[t!]
\centering
\includegraphics[width=0.6\textwidth]{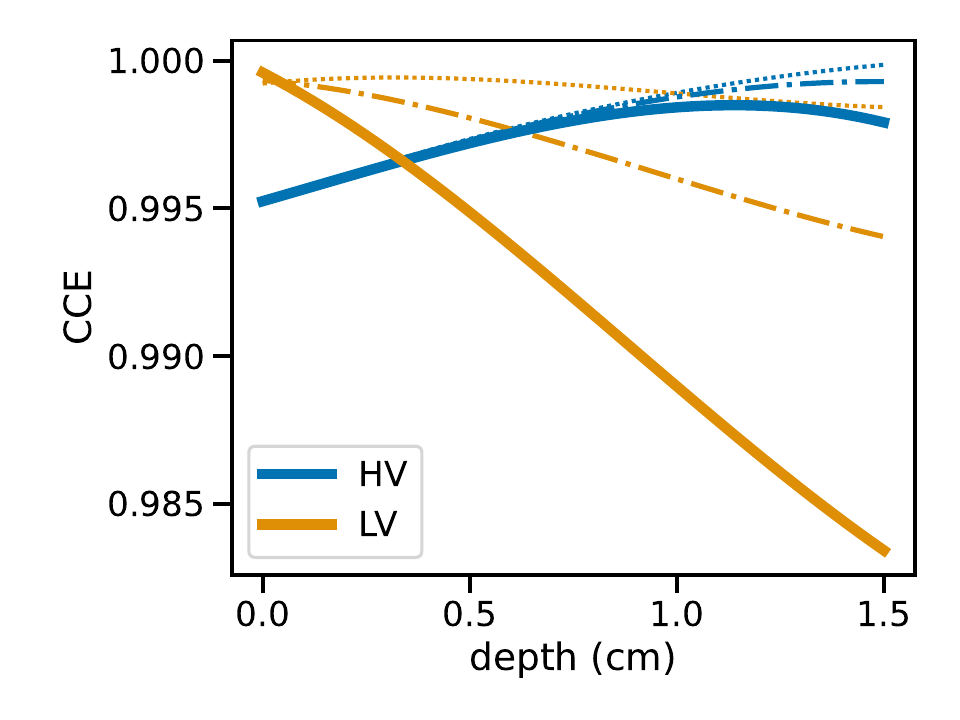}
\caption{: Simulated HV (blue) and LV (orange) electrode charge collection efficiency. The CCEs assuming intrinsic trapping are plotted as dotted lines. The dash-dotted lines represent the expected CCEs given an anticipated proton fluence of $4.2\times10^{7}\,\mathrm{p^{+}\,cm^{-2}}$ during the first two years of the COSI-SMEX mission. The solid lines illustrate the CCEs for a proton fluence of $1.11\times10^{8}\,\mathrm{p^{+}\,cm^{-2}}$, representing the 95\% confidence level upper limit placed on the anticipated fluence in the first two years of the mission.}
\label{fig:predictions}
\end{figure}

We have presented the effects of high-energy proton damage on charge trapping and energy reconstruction in cross-strip GeDs. We found that the detector exhibited intrinsic charge trapping, with electron trapping dominating over hole trapping in the undamaged state. We also showed that hole trapping increased significantly due to proton damage, and we directly characterized the relationship between hole trap density and proton fluence. After applying a proton fluence of $2\times10^{8}\,\mathrm{p^{+}\,cm^{-2}}$, we found that the hole trapping product, $[n\sigma]^{-1}_\mathrm{h}$, decreased from $1786\pm31$\,cm to $106.2\pm0.5$\,cm. At a fluence of $4.95\times10^{8}\,\mathrm{p^{+}\,cm^{-2}}$, the hole trapping product had further decreased to $36.9\pm0.4$\,cm. Using these measurements, we inferred a linear relationship between the hole trap density and the proton fluence, and calculated the rate at which cross-sectional hole trapping area is generated by proton bombardment. This quantitative relationship allows us to predict the amount of hole trapping that will be induced in the COSI-SMEX detectors while they are in space. 

Based on estimates from the SPENVIS package \citep{SPENVIS} for a 530\,km, $0^\circ$ inclination orbit, we anticipate\footnote{We assumed a 50\,MeV threshold for protons to enter the aperture of the instrument and penetrate the germanium and a 150\,MeV threshold for protons to pass through the active BGO shielding and penetrate the germanium from the sides and bottom of the instrument.} a proton fluence of $4.2\times10^{7}\,\mathrm{p^{+}\,cm^{-2}}$ during the first two years of the mission lifetime, with a 95\% confidence level upper limit of $1.11\times10^{8}\,\mathrm{p^{+}\,cm^{-2}}$. For a detector with the same intrinsic hole trapping as the GeD we have investigated, these values correspond to hole trapping products of $[n\sigma]^{-1}_\mathrm{h}=475$\,cm and $[n\sigma]^{-1}_\mathrm{h}=170$\,cm, respectively. We plot the resulting CCEs in Figure \ref{fig:predictions}, with the result for the expected fluence shown as dash-dotted lines and the result for the 95\% upper limit on proton fluence shown as the solid lines. While this is less hole trapping than we have produced in our experiment, it is still significantly greater than the expected intrinsic hole trapping in an undamaged detector and will result in decreased spectral resolution, highlighting the necessity of correcting for trapping.

By using a relatively simple second-order, depth-dependent energy correction, we were able to significantly improve the FWHM and FWTM of spectral lines measured using both the HV and LV strip signals. However, we were unable to completely compensate for the effects of hole trapping. Notably, the LV signal achieved before the detector was damaged demonstrated superior resolution compared to the HV signal both before and after applying trapping corrections. This fact, as well as the large improvements achieved for the damaged detector, demonstrate the importance that measurements of and corrections for electron and hole trapping will play during the COSI mission and for similar missions. As we further improve our trapping correction technique, for example by improving the precision of our depth calibration, we aim to achieve spectral resolution comparable to or better than those obtained prior to proton damage.

\backmatter

\bmhead{Acknowledgments}

This work was supported by the NASA Astrophysics Research and Analysis (APRA) program, grant 80NSSC22K1881, and by the Laboratory Directed Research and Development Program of Lawrence Berkeley National Laboratory under U.S. Department of Energy Contract No. DE-AC02-05CH11231.
We would like to acknowledge the staff at the James M. Slater, MD, Proton Treatment and Research Center, Loma Linda University Medical Center beamline for their assistance in this study and for the precision with which they were able to achieve the desired proton fluence.
We thank the anonymous reviewer for comments and suggestions which improved the quality of this manuscript.

\bibliography{main}

\end{document}